\documentclass{llncs}
\usepackage{a4wide,pst-all}
\usepackage{pifont}
\newcommand{\cache}{\raisebox{-0.5ex}{\mbox{\Large\ding{41}}}\,}
\def\itp#1{(\textit{#1}\/)}
\newsavebox{\boxtabbing}
{\end{tabbing}\end{minipage}\end{lrbox}%
\framebox[\columnwidth][l]{\usebox{\boxtabbing}}}
\newcommand{\remove}[1]{}
\begin{document}

\title{A Distributed TDMA Slot Assignment Algorithm for Wireless
Sensor Networks}

\author{Ted Herman\inst{1} \and S\'{e}bastien Tixeuil\inst{2}}
\institute{
University of Iowa, \email{ted-herman@uiowa.edu} 
\and
LRI -- CNRS UMR 8623 \& INRIA Grand Large,
Universit\'{e} Paris-Sud XI, France, \email{tixeuil@lri.fr}
}

\maketitle

\begin{abstract}
Wireless sensor networks benefit from communication protocols
that reduce power requirements by avoiding frame collision.  
Time Division Media Access methods schedule transmission in 
slots to avoid collision, however these methods often lack  
scalability when implemented in \emph{ad hoc} networks subject to 
node failures and dynamic topology.  This paper reports a 
distributed algorithm for TDMA slot assignment that is self-stabilizing
to transient faults and dynamic topology change.  The expected 
local convergence time is $O(1)$ for any size network satisfying a 
constant bound on the size of a node neighborhood. 
\end{abstract}

\section{Introduction}

Collision management and avoidance are fundamental issues in wireless
network protocols.   Networks now being imagined for sensors \cite{ZG03} and
small devices \cite{CHBSW01} require energy conservation, scalability, 
tolerance to transient faults, and adaptivity to topology change.
Time Division Media Access (TDMA) is a reasonable 
technique for managing wireless media access, however the priorities of 
scalability and fault tolerance are not emphasized by most 
previous research.  Recent analysis \cite{HL03} of radio transmission 
characteristics typical of sensor networks shows that TDMA may not
substantially improve bandwidth when compared to randomized collision
avoidance protocols, however fairness and energy conservation considerations
remain important motivations.  In applications with predictable 
communication patterns, a sensor may even power off the radio receiver 
during TDMA slots where no messages are expected;  such timed approaches
to power management are typical of the sensor regime.

Emerging models of \emph{ad hoc} sensor networks are more constrained than 
general models of distributed systems, especially with respect to 
computational and communication resources.  These constraints tend
to favor simple algorithms that use limited memory.  A few constraints
of some sensor networks can be helpful:  sensors may have access 
to geographic coordinates and a time base (such as GPS provides), 
and the density of sensors in an area can have a known, fixed upper bound.
The question we ask in this paper is how systems can distributively
obtain a TDMA assignment of slots to nodes, given the assumptions of 
synchronized clocks and a bounded density (where density is interpreted
to be a fixed upper bound on the number of immediate neighbors in the
communication range of any node).  In practice, such a limit on the
number of neighbors in range of a node has been achieved by dynamically
attenuating transmission power on radios.  Our answers to the question
of distributively obtaining a TDMA schedule are partial:  our results
are not necessarily optimum, and although the algorithms we present
are self-stabilizing, they are not optimally designed for all cases
of minor disruptions or changes to a stabilized sensor network.  

Before presenting our results, it may be helpful for the 
reader to consider the relation between TDMA scheduling and 
standard problems of graph coloring (since these topics often found
in textbooks on network algorithms for spatial multiplexing).
Algorithmic research on TDMA relates the problem of timeslot 
assignment to minimal graph coloring where the coloring constraint is 
typically that of ensuring that no two nodes within distance
two have the same color (the constraint of distance two has a 
motivation akin to the well known hidden terminal problem in 
wireless networks).  This simple reduction of TDMA timeslot
assignment neglects some opportunities for time division: 
even a solution to minimum coloring does not necessarily
give the best result for TDMA slot assignment.
Consider the two colorings shown in Figure \ref{fig:solution-A}, 
which are minimum distance-two colorings of the same network.
We can count, for each node $p$, the size of the set of colors used
within its distance-two neighborhood (where this set includes 
$p$'s color); this is illustrated in Figure \ref{fig:solution-B} for
the respective colorings of Figure \ref{fig:solution-A}.  We see
that some of the nodes find more colors in their distance-two 
neighborhoods in the second coloring of Figure \ref{fig:solution-A}.
The method of slot allocation in Section \ref{sec:slots}
allocates larger bandwidth share when the number of colors
in distance-two neighborhoods is smaller.
Intuitively, if some node $p$ sees $k<\lambda$ colors in its 
distance-two neighborhood, then it should have at least a $1/(k+1)$ 
share of bandwidth, which is superior to assigning a $1/(\lambda+1)$
share to each color.    
Thus the problem of optimum TDMA slot assignment is, in some 
sense, harder than optimizing the global number of colors.

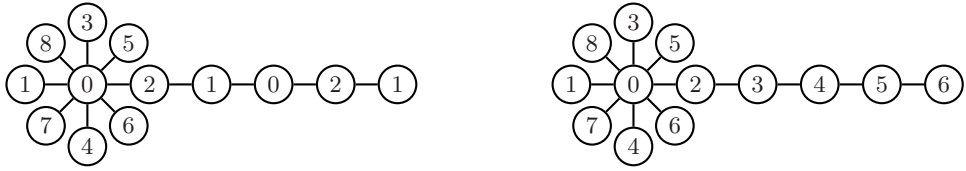
\begin{figure}[t]
\begin{tabular}{cc}
\begin{minipage}{0.48\columnwidth}
\psset{unit=0.55cm}
\begin{pspicture}(-3,-0.5)(4,4)
\rput(2,2){\circlenode{A}{\footnotesize 0}}
\rput(0.5,2){\circlenode{B}{\footnotesize 1}}
\rput(3.5,2){\circlenode{C}{\footnotesize 2}}
\rput(2,3.5){\circlenode{D}{\footnotesize 3}}
\rput(2,0.5){\circlenode{E}{\footnotesize 4}}
\rput(3,3){\circlenode{V}{\footnotesize 5}}\ncline{A}{V}
\rput(3,1){\circlenode{W}{\footnotesize 6}}\ncline{A}{W}
\rput(1,1){\circlenode{X}{\footnotesize 7}}\ncline{A}{X}
\rput(1,3){\circlenode{Y}{\footnotesize 8}}\ncline{A}{Y}
\rput(5,2){\circlenode{F}{\footnotesize 1}}
\rput(6.5,2){\circlenode{G}{\footnotesize 0}}
\rput(8,2){\circlenode{H}{\footnotesize 2}}
\rput(9.5,2){\circlenode{I}{\footnotesize 1}}
\ncline{A}{B}\ncline{A}{C}\ncline{A}{D}\ncline{A}{E}
\ncline{C}{F}\ncline{F}{G}\ncline{G}{H}\ncline{H}{I}
\end{pspicture}
\end{minipage}
&
\begin{minipage}{0.48\columnwidth}
\psset{unit=0.55cm}
\begin{pspicture}(-3,-0.5)(5,4)
\rput(2,2){\circlenode{A}{\footnotesize 0}}
\rput(0.5,2){\circlenode{B}{\footnotesize 1}}
\rput(3.5,2){\circlenode{C}{\footnotesize 2}}
\rput(2,3.5){\circlenode{D}{\footnotesize 3}}
\rput(2,0.5){\circlenode{E}{\footnotesize 4}}
\rput(3,3){\circlenode{V}{\footnotesize 5}}\ncline{A}{V}
\rput(3,1){\circlenode{W}{\footnotesize 6}}\ncline{A}{W}
\rput(1,1){\circlenode{X}{\footnotesize 7}}\ncline{A}{X}
\rput(1,3){\circlenode{Y}{\footnotesize 8}}\ncline{A}{Y}
\rput(5,2){\circlenode{F}{\footnotesize 3}}
\rput(6.5,2){\circlenode{G}{\footnotesize 4}}
\rput(8,2){\circlenode{H}{\footnotesize 5}}
\rput(9.5,2){\circlenode{I}{\footnotesize 6}}
\ncline{A}{B}\ncline{A}{C}\ncline{A}{D}\ncline{A}{E}
\ncline{C}{F}\ncline{F}{G}\ncline{G}{H}\ncline{H}{I}
\end{pspicture}
\end{minipage}
\end{tabular}
\caption{two solutions to distance-two coloring}
\label{fig:solution-A}
\end{figure}
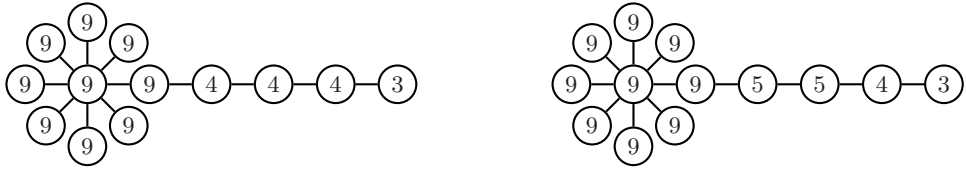
\begin{figure}[hbt]
\begin{tabular}{cc}
\begin{minipage}{0.48\columnwidth}
\psset{unit=0.55cm}
\begin{pspicture}(-3,-0.5)(5,4)
\rput(2,2){\circlenode{A}{\footnotesize 9}}
\rput(0.5,2){\circlenode{B}{\footnotesize 9}}
\rput(3.5,2){\circlenode{C}{\footnotesize 9}}
\rput(2,3.5){\circlenode{D}{\footnotesize 9}}
\rput(2,0.5){\circlenode{E}{\footnotesize 9}}
\rput(3,3){\circlenode{V}{\footnotesize 9}}\ncline{A}{V}
\rput(3,1){\circlenode{W}{\footnotesize 9}}\ncline{A}{W}
\rput(1,1){\circlenode{X}{\footnotesize 9}}\ncline{A}{X}
\rput(1,3){\circlenode{Y}{\footnotesize 9}}\ncline{A}{Y}
\rput(5,2){\circlenode{F}{\footnotesize 4}}
\rput(6.5,2){\circlenode{G}{\footnotesize 4}}
\rput(8,2){\circlenode{H}{\footnotesize 4}}
\rput(9.5,2){\circlenode{I}{\footnotesize 3}}
\ncline{A}{B}\ncline{A}{C}\ncline{A}{D}\ncline{A}{E}
\ncline{C}{F}\ncline{F}{G}\ncline{G}{H}\ncline{H}{I}
\end{pspicture}
\end{minipage}
&
\begin{minipage}{0.48\columnwidth}
\psset{unit=0.55cm}
\begin{pspicture}(-3,-0.5)(5,4)
\rput(2,2){\circlenode{A}{\footnotesize 9}}
\rput(0.5,2){\circlenode{B}{\footnotesize 9}}
\rput(3.5,2){\circlenode{C}{\footnotesize 9}}
\rput(2,3.5){\circlenode{D}{\footnotesize 9}}
\rput(2,0.5){\circlenode{E}{\footnotesize 9}}
\rput(3,3){\circlenode{V}{\footnotesize 9}}\ncline{A}{V}
\rput(3,1){\circlenode{W}{\footnotesize 9}}\ncline{A}{W}
\rput(1,1){\circlenode{X}{\footnotesize 9}}\ncline{A}{X}
\rput(1,3){\circlenode{Y}{\footnotesize 9}}\ncline{A}{Y}
\rput(5,2){\circlenode{F}{\footnotesize 5}}
\rput(6.5,2){\circlenode{G}{\footnotesize 5}}
\rput(8,2){\circlenode{H}{\footnotesize 4}}
\rput(9.5,2){\circlenode{I}{\footnotesize 3}}
\ncline{A}{B}\ncline{A}{C}\ncline{A}{D}\ncline{A}{E}
\ncline{C}{F}\ncline{F}{G}\ncline{G}{H}\ncline{H}{I}
\end{pspicture}
\end{minipage}
\end{tabular}
\caption{number of colors used within distance two}
\label{fig:solution-B}
\end{figure}

\textbf{Contributions.}  The main issues for our research are
dynamic network configurations, transient fault tolerance and scalability 
of TDMA slot assignment algorithms.   Our approach to both dynamic 
network change and transient fault events is to use the paradigm of
self-stabilization, which ensures the system state converges to a 
valid TDMA assignment after any transient fault or topology change event.
Our approach to scalability is to propose a randomized 
slot assignment algorithm with $O(1)$ expected \emph{local} convergence time.  
The basis for our algorithm is, in essence, a probabilistically 
fast clustering technique (which could be exploited for other 
problems of sensor networks).  The expected time for \emph{all} nodes
to have a valid TDMA assignment is not $O(1)$; our view 
is that stabilization over the entire network is an unreasonable 
metric for sensor network applications;  we discuss this further
in the paper's conclusion.  Our approach guarantees that 
after stabilization, if nodes crash, TDMA collision may occur only 
locally (in the distance-three neighborhood of the faults).   
 
\textbf{Related Work.}  The idea of self-stabilizing TDMA has been
developed in \cite{KA03a,KA03b} for model that is more restricted
than ours (a grid topology where each node knows its location).
Algorithms for allocating TDMA time slots and FDMA frequencies
are formulated as vertex coloring problems in a graph \cite{Ram99}.  
Let the set of vertex colors be the integers from the range $0..\lambda$.  
For FDMA the colors $(f_v,f_w)$ of neighboring vertices $(v,w)$ should 
satisfy $|f_v-f_w|>1$ to avoid interference.  The standard 
notation for this constraint is $L(\ell_1,\ell_2)$:   for any 
pair of vertices at distance $i\in\{1,2\}$, the colors differ
by at least $\ell_i$.  The coloring problem for TDMA is:
let $L'(\ell_1,\ell_2)$ be the constraint that for any pair 
of vertices at distance $i\in\{1,2\}$, the colors differ by at 
least $\ell_i$ mod ($\lambda+1)$.
(This constraint represents the fact that 
time slots wrap around, unlike frequencies.)  
The coloring constraint for TDMA is $L'(1,1)$.  
Coloring problems with constraints $L(1,0)$, $L(0,1)$, $L(1,1)$,
and $L(2,1)$ have been well-studied not only for general graphs
but for many special types of graphs \cite{BKTL00,KMR98,KL93};
many such problems are NP-complete and although 
approximation algorithms have been 
proposed, such algorithms are typically not distributed.  
(The related problem finding a minimum dominating 
set has been shown to have a distributed approximation using
constant time \cite{KW03}, though it is unclear if the techniques
apply to self-stabilizing coloring.)  Self-stabilizing algorithms
for $L(1,0)$ have been studied in \cite{GK93,SS93,SRR94,SRR95,GT00},
and for $L(1,1)$ in \cite{GJ01}.
Our algorithms borrow from techniques of self-stabilizing
coloring and renaming \cite{GJ01,GT00}, which use techniques 
well-known in the literature of parallel algorithms on 
PRAM models \cite{Lub86}.  To the extent that the 
sensor network model is synchronous, some of these techniques 
can be adapted; however working out details when messages 
collide, and the initial state is unknown, is not an entirely trivial task.  
This paper is novel in the
sense that it composes self-stabilizing algorithms for renaming and
coloring for a base model that has only probabilistically correct 
communication, due to the possibility of collisions at the media access layer.
Also, our coloring uses a constant number of colors for the
$L(1,1)$ problem, while the previous self-stabilizing solution
to this problem uses $n^2$ colors.

\section{Wireless Network, Program Notation} \label{sec:model}

The system is comprised of a set $V$ of nodes in an \emph{ad hoc} wireless 
network, and each node has a unique identifier.  
Communication between nodes uses a low-power radio.
Each node $p$ can communicate with a subset $N_p\subseteq V$ of nodes
determined by the range of the radio signal;  $N_p$ is called the 
neighborhood of node $p$.  In the wireless model,
transmission is omnidirectional:  each message sent by $p$ is 
effectively broadcast to all nodes in $N_p$.  We also assume that 
communication capability is bidirectional:  $q\in N_p$ iff $p\in N_q$.  
Define $N^1_p=N_p$ and for $i>1$, 
$ N^i_p \;  = \; 
N^{i-1}_p \;\cup\; \{\, r\; | \; 
(\exists q: \;  q\in N^{i-1}_p: \; r\in N_q) \, \}$ 
(call $N^i_p$ the distance-$i$ neighborhood of $p$).
Distribution of nodes is sparse: there is some known 
constant $\delta$ such that for any node $p$, $|N_p|\leq\delta$.  
(Sensor networks can control density by powering off nodes in areas
that are too dense, which is one aim of topology control algorithms.)

Each node has fine-grained, real-time clock hardware, and  
all node clocks are synchronized to a common, global time.  
Each node uses the same radio frequency (one frequency is shared 
spatially by all nodes in the network) and media access is 
managed by CSMA/CA: if node $p$ has a
message ready to transmit, but is receiving some signal, then 
$p$ does not begin transmission until it detects the absence of 
signal; and before $p$ transmits a message, it waits for 
some random period (as implemented, for instance, in \cite{WD01}).
We assume that the implementation of CSMA/CA satisfies the 
following:  there exists a constant 
$\tau>0$ such that the probability of a frame transmission
without collision is at least $\tau$ (this corresponds to 
typical assumptions for multiaccess channels \cite{BG87}; 
the independence of $\tau$ for different frame transmissions
indicates our assumption of an underlying memoryless probability
distribution in a Markov model).

\textbf{Notation.}  We describe algorithms using the
notation of guarded assignment statements:
$G\rightarrow S$ represents a guarded assignment,
where $G$ is a predicate of the local variables of
a node, and $S$ is an assignment to local variables
of the node.  
If predicate $G$ (called the \emph{guard}) holds, 
then assignment $S$ is executed, otherwise $S$ is skipped.
Some guards can be event predicates that hold upon the event of 
receiving a message:  we assume that all such guarded assignments
execute atomically when a message is received.     
At any system state where a given guard $G$ holds, we say
that $G$ is \emph{enabled} at that state.  The $[]$ operator 
is the nondeterministic composition of guarded assignments;  
$([]q: \; q\in M_p: \; G_q\rightarrow S_q)$ is a closed-form expression of   
$G_{q1}\rightarrow S_{q_1} \;[]\; G_{q_2}\rightarrow S_{q_2} 
	\;[]\; \cdots
   \;[]\; G_{q_k}\rightarrow S_{q_k}$, where $M_p=\{q_1,q_2,\ldots,q_k\}$. 

\textbf{Execution Semantics.}  The life of computing at every
node consists of the infinite repetition of 
finding a guard and executing its corresponding assignment or 
skipping the assignment if the guard is \emph{false}. 
Generally, we suppose that when a node executes its program,
all statements with \emph{true} guards are executed 
in some constant time (done, for example, in round-robin order).

\paragraph{Shared Variable Propagation}

A certain subset of the variables at any node are 
designated as \emph{shared} variables.
Nodes periodically transmit the values of their shared variables,
based on a timed discipline.  A simple protocol in our notation for
periodic retransmission would be $\textit{true}\rightarrow 
\textit{transmit}(\textit{var}_p)$ for each shared variable of $p$,
where generally, we suppose that when a node executes its program,
all statements with \emph{true} guards are executed
in some constant time (done, for example, in round-robin order).
(One local node variable we do not 
explicitly use is the clock, which advances continuously in real time;  
guards and assignments could refer to the clock, but we prefer to 
discipline the use of time as follows.)  

Beyond periodic retransmission, assignment to a shared variable
causes peremptory transmission: if a statement $G\rightarrow S$ 
assigns to a shared variable, then we present the 
statement without any reference to the clock and we suppose that
there is a transformation of the statement into a computation that 
slows execution so that it does not exceed some desired rate, 
and also provides randomization to avoid collision in messages 
that carry shared variable values.
This could be implemented using 
a timer associated with $G\rightarrow S$.  One technique for 
implementing $G\rightarrow S$ could be by the following procedure
\begin{quote}
Suppose the previous invocation of the procedure for 
$G\rightarrow S$ finished at time $t$;  the next evaluation 
of $G\rightarrow S$ occurs at time $t+\beta$, where $\beta$ is 
a random delay inserted by the CSMA/CA implementation.
After executing $S$ (or skipping the assignment 
if $G$ is \textit{false}), the node
transmits a message containing all shared variable values.
This message transmission may be postponed if the node is
currently receiving a message.  Finally, after transmitting the message,
the node waits for an additional $\kappa$ time units, where $\kappa$
is a given constant.  Thus, in brief, $G\rightarrow S$ is forever 
evaluated by waiting for a random period, atomically evaluating
$G\rightarrow S$, transmitting shared variable(s), 
and waiting for a constant time period $\kappa$.
Figure \ref{fig:random-send} illustrates the cycle of shared
variable propagation for one node. 
\end{quote}
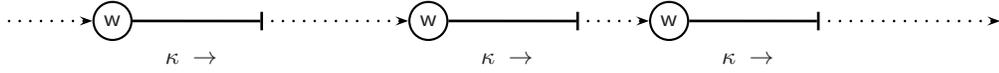
\begin{figure}[t]
\begin{pspicture}(-3,0)(5,2)
%\psset{unit=0.5cm}
\pnode(-1.5,1){I}
\rput(0,1){\circlenode{A}{\footnotesize \textsf{w}}}
\rput(1,0.5){\footnotesize $\;\kappa\;\rightarrow$}
\pnode(2,1){B}
\rput(4.2,1){\circlenode{C}{\footnotesize \textsf{w}}}
\rput(5.2,0.5){\footnotesize $\;\kappa\;\rightarrow$}
\pnode(6.2,1){D}
\rput(7.4,1){\circlenode{E}{\footnotesize \textsf{w}}}
\rput(8.4,0.5){\footnotesize $\;\kappa\;\rightarrow$}
\pnode(9.4,1){F}
\pnode(11.8,1){G}
\psset{linewidth=1pt}
\ncline[linestyle=dotted]{->}{I}{A}
\ncline{-|}{A}{B}
\ncline[linestyle=dotted]{->}{B}{C}
\ncline{-|}{C}{D}
\ncline[linestyle=dotted]{->}{D}{E}
\ncline{-|}{E}{F}
\ncline[linestyle=dotted]{->}{F}{G}
\end{pspicture}
\caption{shared variable propagation}
\label{fig:random-send}
\end{figure}

To reconcile our earlier assumption of immediate, atomic processing
of messages with the discipline of shared variable assignment, no
guarded assignment execution should change a shared variable 
in the atomic processing of receiving a message.  All the programs in 
this paper have this property, that receipt of a message atomically
changes only nonshared variables.  

Given the discipline of repeated transmission of 
shared variables, each node can have a cached copy of the value
of a shared variable for any neighbor.  This cached copy is 
updated atomically upon receipt of a message carrying a new
value for the shared variable.  

\paragraph{Model Construction}

Our goal is to provide an implementation of a general purpose, 
collision-free communication service.   This service can be 
regarded as a transformation of the given model of Section
\ref{sec:model} into a model without collisions.  This service
simplifies application programming and can reduce energy 
requirements for communication (messages do not need
to be resent due to collisions).  Let $\mathcal T$ denote
the task of transforming the model of Section \ref{sec:model}
into a collision-free model.

We seek a solution to $\mathcal T$ that is self-stabilizing
in the sense that, after some transient failure or reconfiguration,
node states may not be consistent with the requirements of 
collision-free communication and collisions can occur; eventually
the transformer corrects node states to result in collision-free
communication.  Our first design decision is to suppose that the
implementation we seek is not itself free of collisions.
That is, even though our goal is to provide applications a 
collision-free service, our implementation may introduce 
overhead messages susceptible to collisions.  Initially,
in the development of algorithms, we accept 
collisions and resends of these messages, which are 
internal to $\mathcal T$ and not visible to the application.  

To solve $\cal T$ it suffices to assign each 
node a color and use node colors as the schedule for a TDMA
approach to collision-free communication \cite{Ram99}.  
Even before colors are assigned, we use a schedule that partitions radio
time into two parts: one part is for TDMA scheduling of application
messages and the other part is reserved for the messages
of the algorithm that assigns colors and time slots to nodes.
The following diagram illustrates such a schedule,
in which each TDMA part has five slots.  Each overhead part is,
in fact, a fixed-length slot in the TDMA schedule.

{\footnotesize
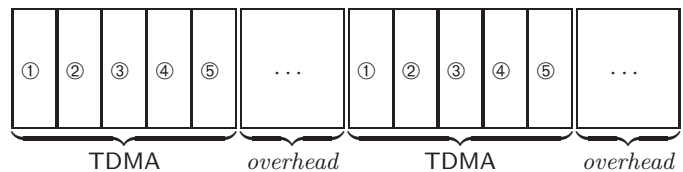
   \[
   \underbrace{
   \fbox{\raisebox{4ex}{\ding{192}\raisebox{6ex}{~}}}
   \fbox{\raisebox{4ex}{\ding{193}\raisebox{6ex}{~}}}
   \fbox{\raisebox{4ex}{\ding{194}\raisebox{6ex}{~}}}
   \fbox{\raisebox{4ex}{\ding{195}\raisebox{6ex}{~}}}
   \fbox{\raisebox{4ex}{\ding{196}\raisebox{6ex}{~}}}
   }_{\textsf{TDMA}} 
   \underbrace{
   \fbox{\raisebox{4ex}{\quad$\cdots$\raisebox{6ex}{~}\quad}}
   }_{\textit{\small overhead}}
   \underbrace{
   \fbox{\raisebox{4ex}{\ding{192}\raisebox{6ex}{~}}}
   \fbox{\raisebox{4ex}{\ding{193}\raisebox{6ex}{~}}}
   \fbox{\raisebox{4ex}{\ding{194}\raisebox{6ex}{~}}}
   \fbox{\raisebox{4ex}{\ding{195}\raisebox{6ex}{~}}}
   \fbox{\raisebox{4ex}{\ding{196}\raisebox{6ex}{~}}}
   }_{\textsf{TDMA}} 
   \underbrace{
   \fbox{\raisebox{4ex}{\quad$\cdots$\raisebox{6ex}{~}\quad}}
   }_{\textit{\small overhead}}
   \]
}

The programming model, including the technique for sharing 
variables described in Section \ref{sec:model}, refers to message
and computation activity in the overhead parts.  It should be
understood that the timing of shared variable propagation
illustrated in Figure \ref{fig:random-send} may span overhead
slots:  the computation by the solution to $\cal T$ operates
in the concatenation of all the overhead slots.  Whereas CSMA/CA 
is used to manage collisions in the overhead slots, the remaining
TDMA slots do not use random delay.  During initialization or 
after a dynamic topology change, frames may collide in the 
TDMA slots, but after the slot assignment algorithm self-stabilizes,
collisions do not occur in the TDMA slots.

With respect to any given node $v$, a solution $\mathcal T$ is 
\emph{locally stabilizing} with convergence time $t$ if, 
for any initial system state, after at most $t$ time units,
every subsequent system state satisfies the property that 
any transmission by $v$ during its assigned slot(s) is 
free from collision.  Solution $\mathcal T$ is \emph{globally
stabilizing} with convergence time $t$ if, for every initial
state, after at most $t$ time units, every subsequent system
state has the property that all transmissions during assigned
slots are free from collision.  For randomized algorithms, 
these definitions are modified to specify expected convergence
times (all stabilizing randomized algorithms we consider are
probabilistically convergent in the Las Vegas sense).  When
the qualification (local or global) is omitted, convergence
times for local stabilization are intended for the presented 
algorithms.

Several primitive services that are not part of the initial
model can simplify the design and expression of $\mathcal T$'s
implementation.  All of these services need to be self-stabilizing. 
Briefly put, our plan is to develop a sequence of algorithms that
enable TDMA implementation.  These algorithms are:  
neighborhood-unique naming, maximal independent set, 
minimal coloring, and the assignment of time slots from colors.
In addition, we rely on neighborhood
services that update cached copies of shared variables. 

\paragraph{Neighborhood Identification}
\label{sec:identification}

We do not assume that a node $p$ has built-in knowledge of 
its neighborhood $N_p$
or its distance-three neighborhood $N^3_p$.  This is because the
type of network under considering is \emph{ad hoc}, and the 
topology dynamic.  Therefore some algorithm is needed so that a
node can refer to its neighbors.  We describe first how a node $p$
can learn of $N^2_p$, since the technique can be extended to learn
$N^3_p$ in a straightforward way.

Each node $p$ can represent $N^i_p$ for $i\in 1..3$ by a list of 
identifiers learned from messages received at $p$.   However, because
we do not make assumptions about the initial state of any node, such
list representations can initially have arbitrary data.   
Let $L$ be a data type for a list of up to $\delta$ items of the form 
$a:A$, where $a$ is an identifier and $A$ is a set of up to 
$\delta$ identifiers.  Let $sL_p$ be a shared variable of type $L$.  
Let message type \textit{mN} with field of type $L$
be the form of messages transmitted for $sL_p$.
Let $L_p$ be a private variable of a type that is an 
augmentation of $L$ -- it associates a real number
with each item: $\textit{age}(a:A)$ is a positive real value  
attached to the item.

Function $\textsf{update}(L_p,a:A)$
changes $L_p$ to have new item information:  
if $L_p$ already has some item whose first component is $a$, 
it is removed and replaced with $a:A$ (which then has age zero);
if $L_p$ has fewer than $\delta$ items and no item with
$a$ as first component, then $a:A$ is added to $L_p$;
if $L_p$ has already $\delta$ items and no item with
$a$ as first component, then $a:A$ replaces some item with 
maximal age.

Let \textsf{maxAge} be some constant designed to be an upper 
limit on the possible age of items in $L_p$.  Function   
$\textsf{neighbors}(L_p)$ returns the set 
\[ \{\,q\; | \; q\neq p \; \wedge\; 
(\exists\,(a:A): \; (a:A)\in L_p: \; a=q)\; \} \]
Given these variable definitions and functions, we present
the algorithm for neighborhood identification.
\par
\begin{tabbing}
xx \= xxx \= xxx \= \kill
\> \textsf{N0:} \> $\textit{receive}\;\textit{mN}(a:A)\;\rightarrow\;
\textsf{update}(L_p,a:A\setminus\{p\})$ \\
\> \textsf{N1:} \> $([]\,(a:A)\in L_p: \; 
\textit{age}(a:A)>\textsf{maxAge} \;\rightarrow\;$ $L_p := L_p \setminus (a:A)\;)$ \\
\> \textsf{N2:} \> $\textit{true}\;\rightarrow\; sL_p := 
(p:\textsf{neighbors}(L_p))$ 
\end{tabbing}
We cannot directly prove that this algorithm stabilizes
because the CSMA/CA model admits the possibility that a frame, even if 
repeatedly sent, can suffer arbitrarily many collisions.  
Therefore the \textit{age} associated with any element of 
$L_p$ can exceed \textsf{maxAge}, and the element will be 
removed from $L_p$.  The constant \textsf{maxAge} should be
tuned to safely remove old or invalid neighbor data, yet 
to retain current neighbor information by receiving new 
\textit{mN} messages before age expiration.  This is an 
implementation issue beyond of the scope of this paper:  
our abstraction of the behavior of the communication layer
is the assumption that, eventually for any node, the guard 
of \textsf{N1} remains \textit{false} for any $(a:A)\in L_p$ 
for which $a\in N_p$.  
\begin{proposition}
Eventually, for every node $p$, $sL_p=N_p$ 
holds continuously.
\end{proposition}
\begin{proof}
Eventually any element $(a:A)\in L_p$ such that $a\not\in N_p$
is removed.  Therefore, eventually every node $p$ 
can have only its neighbors listed in $sL_p$.  Similarly, 
with probability 1, each node $p$ eventually receives an 
\textit{mN} message from each neighbor, so $sL_p$ contains
exactly the neighbors of $p$. 
\end{proof}
By a similar argument, eventually each node $p$ correctly 
has knowledge of $N^2_p$ as well as $N_p$.  The same technique
can enable each node to eventually have knowledge of $N^3_p$
(it is likely that $N^3_p$ is not necessary; we discuss
this issue in Sections \ref{unique-naming} and \ref{sec:leader}).
In all subsequent sections, we use $N^i_p$ for $i\in 1..3$
as constants in programs with the understanding that such neighborhood 
identification is actually obtained by the stabilizing
protocol described above.

Building upon $L_p$, cached values of the shared variables of
nodes in $N^i_p$, for $i\in 1..3$, can be maintained at $p$;  
erroneous cache values not associated with any node can be discarded by
the aging technique.  We use the following notation in the rest of the
paper:  for node $p$ and some shared variable $\textit{var}_q$
of node $q\in N^3_p$, let $\cache\textit{var}_q$ refer to the 
cached copy of $\textit{var}_q$ at $p$.   The method of 
propagating cached copies of shared variables is generally 
self-stabilizing only for shared variables that do not change
value.  With the exception of one algorithm presented in 
Section \ref{unique-naming}, all of our algorithms use cached
shared variables in this way:  eventually, the shared variables
become constant, implying that eventually all cached copies of
them will be coherent.   
  
For algorithms developed in subsequent sections,
we require a stronger property than eventual propagation of 
shared variable values to their caches.  We require that 
with some constant probability, any shared variable will be
propagated to its cached locations within constant time.  This
is tantamount to requiring that with constant probability, 
a node will transmit within constant time and the transmission
will not collide with any other frame.  Section \ref{sec:model}
states our assumption on wireless transmission, based on the 
constant $\tau$ for collision-free transmission.  The discipline
of shared variable propagation illustrated in Figure 
\ref{fig:random-send} spaces shared-variable updates 
(or skipping updates when there is nothing to change) 
by $\kappa+\beta$, where $\beta$ is a random variable.  Our
requirement on the behavior of transmitting shared variable
values thus also implies that for any time $t$ between 
transmission events, there is a constant probability that
the next transmission event will occur by $t+\alpha$ for 
some constant $\alpha$.  Notice that the joint probability
of waiting at most time $\alpha$, and then sending without
collision, is bounded below by a constant.  It follows that
the expected number of attempts to propagate a shared variable
value before successfully writing to all its caches is $O(1)$.
(In fact, it would not change our analysis if random variable
$\beta$ is truncated by aborting attempted transmissions that
exceed some constant timeout threshold.)  We henceforth assume 
that the expected time for shared variable propagation is constant. 

\paragraph{Problem Definition.}  Let $\mathcal T$ denote
the task of assigning TDMA slots so that each node has some
assigned slot(s) for transmission, and this transmission is 
guaranteed to be collision-free.  We seek a solution to 
$\mathcal T$ that is distributed and self-stabilizing
in the sense that, after some transient failure or reconfiguration,
node states may not be consistent with the requirements of 
collision-free communication and collisions can occur; eventually
the algorithm corrects node states to result in collision-free
communication.

\section{Neighborhood Unique Naming}
\label{unique-naming}

An algorithm providing neighborhood-unique naming gives each
node a name distinct from any of its $N^3$-neighbors.  This may
seem odd considering that we already assume that nodes have 
unique identifiers, but when we try to use the identifiers for
certain applications such as coloring, the potentially large 
namespace of identifiers can cause scalability problems.  Therefore
it can be useful to give nodes smaller names, from a constant space
of names, in a way that ensures names are locally unique.

The problem of neighborhood unique naming can be 
considered as an $N^3$-coloring algorithm and quickly suggests a solution
to $\mathcal T$.  Since neighborhood unique naming provides a solution to
the problem of $L(1,1)$ coloring, it provides a schedule for 
TDMA.  This solution would be especially
wasteful if the space of unique identifiers is larger than 
$|V|$.  It turns out that 
having unique identifiers within a neighborhood
can be exploited by other algorithms to obtain a 
minimal $N^2$-coloring, so we present a simple randomized
algorithm for $N^3$-naming.

Our neighborhood unique naming
algorithm is roughly based on the 
randomized technique described in \cite{GJ01},  
and introduces some new features.
Define $\Delta=\lceil\delta^t\rceil$ for some $t>3$;  the choice 
of $t$ to fix constant $\Delta$ has two competing 
motivations discussed at the end of this section. 
We call $\Delta$ the \emph{namespace}.
Let shared variable $\textit{Id}_p$
have domain $0..\Delta$;  variable $\textit{Id}_p$ is the 
\emph{name} of node $p$.  Another variable is used to collect
the names of neighboring nodes:  
$\textit{Cids}_p=\{\,\cache\textit{Id}_q\;|\;q\in N^3_p\setminus\{p\}\;\}$.
Let $\textsf{random}(S)$ choose with uniform probability
some element of set $S$.  
Node $p$ uses the following function to compute $\textit{Id}_p$: 
\[ \textsf{newId}(\textit{Id}_p) = \left\{ \begin{array}{ll} 
 \textit{Id}_p & \textrm{if} \; \textit{Id}_p\not\in\textit{Cids}_p \\ 
 \textsf{random}(\Delta \setminus \textit{Cids}_p)
 & \textrm{otherwise} \end{array} \right. \]
The algorithm for unique naming is the following.
\begin{tabbing}
xxxxxx \= xxx \= xxx \= \kill
\> \textsf{N3:} \> $\textit{true}\;\rightarrow\; 
\textit{Id}_p := \textsf{newId}(\textit{Id}_p)$ 
\end{tabbing}
Define $\textsf{Uniq}(p)$ to be the predicate that holds iff
\itp{i} no name mentioned in $\textit{Cids}_p$ 
is equal to $\textit{Id}_p$, \itp{ii} for each $q\in N^3_p$, $q\neq p$,
$\textit{Id}_q\neq \textit{Id}_p$, \itp{iii} for each $q\in N^3_p$, 
one name in $\textit{Cids}_q$ equals $\textit{Id}_p$, 
\itp{iv} for each $q\in N^3_p$, $q\neq p$, 
the equality $\cache\textit{Id}_p=\textit{Id}_p$ holds at node $q$,
and \itp{v} no cache update message \emph{en route} to $p$ 
conveys a name that would update $\textit{Cids}_p$ to have 
a name equal to $\textit{Id}_p$.
Predicate $\textsf{Uniq}(p)$ states that $p$'s name is known to 
all nodes in $N^3_p$ and does not conflict with any name of a 
node $q$ within $N^3_q$, nor is there a cached name liable to 
update $\textit{Cids}_p$ that conflicts with $p$'s name.
A key property of the algorithm is the following:  
$\textsf{Uniq}(p)$ is a stable property of the execution.  
This is because after $\textsf{Uniq}(p)$ holds, any node $q$
in $N^3_p$ will not assign $\textit{Id}_q$ to equal $p$'s
name, because \textsf{N3} avoids names listed in the cache
of distance-three neighborhood names -- this stability property
is not present in the randomized algorithm \cite{GJ01}.
The property $(\forall r: \; r\in R: \; \textsf{Uniq}(r))$ 
is similarly stable for any subset $R$ of nodes.  
In words, once a name becomes established as unique
for all the neighborhoods it belongs to, it is stable. 
Therefore we can reason about a Markov model of executions
by showing that the probability of a sequence of steps 
moving, from one stable set of ids to a larger stable set, is positive.

\begin{lemma} \label{lem:uniq} Starting from any state, there is a constant,
positive probability that $\textsf{Uniq}(p)$ holds within
constant time.
\end{lemma}

\begin{proof}
The proof has three cases for $p$: \itp{a} 
$\textsf{Uniq}(p)$ holds initially, 
\itp{b} $\neg\textsf{Uniq}(p)$ holds, but $p$ cannot detect
this locally (this means that there exists some neighbor $q$ of $p$
such that $\cache\textit{Id}_p\neq\textit{Id}_p$ at $q$);
or \itp{c} $p$ detects $\neg\textsf{Uniq}(p)$
and chooses a new name.  Case \itp{a} trivially verifies the lemma.
For case \itp{b}, it could
happen that $\textsf{Uniq}(p)$ is established only by 
actions of nodes other than $p$ within constant time,
and the lemma holds;  otherwise we rely on 
the periodic mechanism of cache propagation and the 
lower bound $\tau$ on the probability of collision-free 
transmission to reduce \itp{b} to \itp{c} with 
some constant probability within constant time.   
For case \itp{c} we require a joint event, which is
the following sequence: $p$ chooses a name
different from any in $N^3_p$ and their caches (or messages
\emph{en route}), then $p$ transmits the new name without 
collision to $N_p$, each node $q\in N_p$ transmits 
the cache of $p$'s name without collision, and then
each node in $N^2_p\setminus N_p$ transmits the cache
of $p$'s name without collision.  Fix some 
constant time $\Phi$ for this sequence of events; time 
$\Phi$ could be $(\delta+1)\cdot\mu$, where $\mu$ is the
average time for a cached value to be transmitted without
collision.  The joint probability $x$ for this scenario is the 
product of probabilities for each event, with the constraint
that the event is transmission without collision within the
desired time constraint $\mu$.  This sequence is not enough,
however to fully estimate the probability for case \itp{c},
because it could be that nodes of $N^3_p$ concurrently 
assign new identifiers, perhaps equal to $p$'s name.  
Therefore we multiply by $x$ the product of probabilities 
that each invocation of \textsf{newId} by $q\in N^3_p$ during 
the time period $\Phi$ does not return a name equal to $p$'s name.
Notice that the number of times that any $q\in N^3_p$
can invoke \textsf{N3} is bounded by $\Phi/\kappa$, because
assignment to shared variables follows the discipline of at
least $\kappa$ delay.  Thus the entire number of invocations
of \textsf{newId} in the $\Phi$-length time period is bounded
by a constant.  Therefore the overall joint probability is
estimated by the product of $x$ and a fixed number of 
constant probabilities;  the joint probability 
for this scenario is thus bounded by a product of constant 
probabilities (dependent on $\Delta$, $\delta$, $\tau$, and $\kappa$).  
Because this joint probability is bounded below by a nonzero constant, 
the expected number of trials to reach a successful result is constant. 
\end{proof}

\begin{corollary} 
The algorithm self-stabilizes with probability 1 and
has constant expected local convergence time.
\end{corollary}
\begin{proof}
The Markov chain for the algorithm has a trapping state for
any $p$ such that $\textsf{Uniq}(p)$ holds.  The stability 
of $\textsf{Uniq}(p)$ for each $p$ separately means that
we can reason about self-stabilization for each node 
independently.  The previous lemma implies that each node
converges to $\textsf{Uniq}(p)$ with probability 1, and 
also implies the constant overall time bound.
\end{proof}
Using the names assigned by \textsf{N3} is a solution to 
$L(1,1)$ coloring, however using 
$\Delta$ colors is not the basis for an efficient
TDMA schedule.  The naming obtained by the algorithm
does have a useful property.  
Let $P$ be a path of $t$ distinct nodes, that is, 
$P = p_1,p_2,\ldots,p_t$.
Define predicate $\textit{Up}(P)$ to hold if 
$\textit{id}_{p_i} < \textit{id}_{p_j}$ for each $i<j$.  
In words, $\textit{Up}(P)$ holds if the names along
the path $P$ increase.  
\begin{lemma} \label{dag-lemma}
Every path $P$ satisfying $\textit{Up}(P)$ has fewer than $\Delta+1$ nodes.
\end{lemma}
\begin{proof}
If a path $P$ satisfying $\textit{Up}(P)$ has $\Delta+1$ nodes, then
some name appears at least twice in the path.  The ordering
on names is transitive, which implies that some name $a$ 
of a node in $P$ satisfies $a<a$, and this contradicts the 
total order on names. 
\end{proof}
This lemma shows that the simple coloring algorithm gives us
a property that node identifiers do not have:  the path length of 
any increasing sequence of names is bounded by a constant.
Henceforth, we suppose that node identifiers have this property,
that is, we treat $N^i_p$ as if the node identifiers
are drawn from the namespace of size $\Delta$.
\par
There are two competing motivations for tuning the parameter
$t$ in $\Delta=\delta^t$.  First, $t$ should be large enough to
ensure that the choice made by \textsf{newId} is unique with 
high probability.  In the worst case, $|N^3_p|=\delta^3+1$, 
and each node's cache can contain $\delta^3$ names, so 
a choosing $t\approx 6$ could be satisfactory.  Generally, 
larger values for $t$ decrease the expected convergence time  
of the neighborhood unique naming algorithm.
On the other hand, smaller values of $t$ will reduce
the constant $\Delta$, which will reduce the convergence time for
algorithms described in subsequent sections.

\section{Leaders via Maximal Independent Set}
\label{sec:maxindep}

Simple distance two coloring algorithms may use a  
number of colors that is wastefully large.  Our
objective is to find an algorithm that uses a reasonable number of
colors and completes, with high probability, in constant time.
We observe that an assignment to satisfy distance two
coloring can be done in constant time given a set of 
neighborhood leader nodes distributed in the network.  
The leaders dictate coloring for nearby nodes.  The coloring 
enabled by this method is minimal (not minimum, which is an NP-hard
problem).  An algorithm selecting a maximal independent set 
is our basis for selecting the leader nodes.

Let each node $p$ have a boolean shared variable $\ell_p$.  In an initial
state, the value of $\ell_p$ is arbitrary.  A legitimate state for
the algorithm satisfies $(\forall p: \; p\in V: \; \mathcal{L}_p)$, where
\begin{eqnarray*}
\mathcal{L}_p 
&\equiv & (\ell_p\Rightarrow (\forall q: \; q\in N_p: \; \neg\ell_q))\\
&\wedge & (\neg\ell_p\Rightarrow (\exists q: \; q\in N_p: \; \ell_q))
\end{eqnarray*}
Thus the algorithm should elect one leader (identified by the 
$\ell$-variable) for each neighborhood.
As in previous sections, $\cache \ell_p$ denotes the cached copy of the
shared variable $\ell_p$.

\begin{tabbing}
x \= xxx \= xxx \= xxx \= \kill
\>\textsf{R1:}\> $(\forall q: \; q\in N_p: \; q>p) \;\rightarrow\; 
	\ell_p := \textit{true}$ \\
\>\textsf{R2:}\> $([]\,q: \; q\in N_p: \; \cache\ell_q \;\wedge\; q<p 
	\;\rightarrow\; \ell_p := \textit{false})$ \\
\>\textsf{R3:}\> $(\exists q: \; q\in N_p: \; q<p) \;\wedge\;
	(\forall q: \; q\in N_p \;\wedge\;$  $(q>p \;\vee\; \neg\cache\ell_q)) \;\rightarrow\; 
	\ell_p := \textit{true}$ 
\end{tabbing}

Although the algorithm does not use randomization, its convergence
technically remains probabilistic because our 
underlying model of communication uses CSMA/CA based on random delay.  The
algorithm's progress is therefore guaranteed with probability 1 
rather than by deterministic means.  

\begin{lemma}
With probability 1 the algorithm \textsf{R1}-\textsf{R3} 
converges to a solution of maximal independent set;  the
convergence time is $O(1)$ if each timed variable
propagation completes in $O(1)$ time.
\end{lemma}
\begin{proof} 
We prove by induction on the namespace that each 
node $p$ stabilizes its 
value of $\ell_p$ within $O(\Delta)$ time.  For the 
base case, consider the 
set $S$ of nodes with locally minimum names, that
is, $(\forall p,q: \; p\in S \; \wedge\; q\in N_p: \; p<q)$.  
Any node $p\in S$ stabilizes in $O(1)$ time
to $\ell_p=\textit{true}$.  The claim follows from the fact 
that guards of \textsf{R2} and \textsf{R3} are \emph{false}, 
whereas the guard of \textsf{R1} is permanently \emph{true}.  
Therefore for the induction step, we can ignore \textsf{R1},
as it is dealt with in the base case.

To complete the induction, suppose that each node $r$ has stabilized
the value of $\ell_r$, where $r\leq k$.  Now consider the situation
of a node $p$ with name $k+1$ (if there is no such node, the induction
is trivially satisfied).  As far as the guards of \textsf{R2} and 
\textsf{R3} are concerned, the value of $\ell_q$ is only 
relevant for a neighbor $q$ with $q<p$, and for any such neighbor,
$\ell_q$ is stable by hypothesis.  Since guards of \textsf{R2} and 
\textsf{R3} are exclusive, it follows that $p$ stabilizes $\ell_p$
and $\cache\ell_p$ is propagated within $O(1)$ time.

Finally, we observe that in any fixed point of the algorithm
\textsf{R1}--\textsf{R3}, no two neighbors are leaders (else 
\textsf{R2} would be enabled for one of them), nor does any 
nonleader find no leader in its neighborhood (else \textsf{R1} or
\textsf{R3} would be enabled).  This establishes that $\mathcal{L}_p$
holds at a fixed point for every $p\in V$.  
The induction terminates with at
most $|\Delta|$ steps, the size of the namespace, and because 
$\Delta$ is a constant, the convergence time is $O(1)$ for this
algorithm.
\end{proof}

\section{Leader Assigned Coloring}
\label{sec:leader}

Our method of distance-two coloring is simple:
colors are assigned by the leader nodes given by maximal 
independent set output.  The following variables are introduced
for each node $p$:
\begin{description}
\item[$color_p$] is a number representing the color for node $p$.
\item[$min\ell_p$] is meaningful only for $p$ such that 
$\neg\ell_p$ holds:  it is intended to satisfy 
\[ \textit{min}\ell_p \; = \; 
  \textrm{min} \; \{ \; q \; | \; q\in N_p \;\wedge\; \cache\ell_q \; \} \]
In words, $\textit{min}\ell_p$ is the smallest id of any neighbor
that is a leader.  Due to the uniqueness of names in $N^3_p$, 
the value $\textit{min}\ell_p$ stabilizes to a unique node.
\item[$spectrum_p$] \hspace*{2em} is a set of pairs $(c,r)$ where 
$c$ is a color and $r$ is an id.  Pertaining only to nonleader
nodes, $\textit{spectrum}_p$ should contain 
$(\textit{color}_p,\textit{min}\ell_p)$ and 
$(\cache\textit{color}_q,\,\cache\textit{min}\ell_q)$ for each $q\in N_p$.
\item[$setcol_p$] ~ is meaningful only for $p$ such that
$\ell_p$ holds.  It is an array of colors indexed by identifier:  
$\textit{setcol}_p[q]$ is $p$'s preferred color for $q\in N_p$.  
We consider $\textit{color}_p$ and $\textit{setcol}_p[p]$ to be
synonyms for the same variable.  In the algorithm we use the
notation $\textit{setcol}_p{[A]:=}B$ to denote the parallel assignment 
of a set of colors $B$ based on a set of indices $A$.  To make
this assignment deterministic, we suppose that $A$ can be represented
by a sorted list for purposes of the assignment;  $B$ is similarly
structured as a list.
\item[$dom_p$] for leader $p$ is computed to be the
nodes to which $p$ can give a preferred color;  these include any
$q\in N_p$ such that $\textit{min}\ell_q=p$.
We say for $q\in\textit{dom}_p$ that $p$ \emph{dominates} $q$. 
\item[$f$] is a function used by each leader $p$ 
to compute a set of unused colors
to assign to the nodes in $\textit{dom}_p$.  The set of \emph{used} 
colors for $p$ is 
\[ \{ \, c \; | \; (\exists\,q,r:\; q\in N_p \;\wedge\;
(c,r)\in\textit{spectrum}_q \;\wedge\; r<p)\;\} \]
That is, used colors with respect to $p$ are those colors in
$N^2_p$ that are already assigned by leaders with smaller 
identifiers than $p$.
The complement of the \emph{used} set is the range of possible colors 
that $p$ may prefer for nodes it dominates.  Let $f$ be the 
function to minimize the number of colors preferred for 
the nodes of $\textit{dom}_p$, ensuring that 
the colors for $\textit{dom}_p$ are distinct,
and assigning smaller color indices (as close to 0 as possible)
preferentially.  Function $f$ returns
a list of colors to match the deterministic list of $\textit{dom}_p$
in the assignment of \textsf{R5}.
\end{description}
\begin{tabbing}
xxxx \= xxx \= xxx \= xxx \= \kill
\textsf{R4:} \> $\ell_p \;\rightarrow\; \textit{dom}_p := \{\,p\,\} \cup 
		\{ \, q \, | \; q\in N_p \wedge 
		\cache\textit{min}\ell_q = p \; \}$ \\
\textsf{R5:} \> $\ell_p \;\rightarrow\; \textit{setcol}_p[\textit{dom}_p] :=   
	f(\, \{ c \; | \; \exists q:$  
	$\; q\in N_p \; \wedge \; r<p \;\wedge\;
		(c,r)\in \cache\textit{spectrum}_q \, \} \, )$ \\
\textsf{R6:} \> $\textit{true} \;\rightarrow\; 
	\textit{min}\ell_p := 
	\textrm{min}\,\{ \; q \; | \; 
	q\in N_p \cup \{p\} \;\wedge\; \cache\ell_q \; \}$ \\
\textsf{R7:} \> $\neg\ell_p \;\rightarrow\; 
	\textit{color}_p := 
		\cache\textit{setcol}_r[p]$, where $r=\textit{min}\ell_p$
	\\
\textsf{R8:} \> $\neg\ell_p \;\rightarrow\;
	\textit{spectrum}_p := (\textit{color}_p,\textit{min}\ell_p)
	\;\cup\; \bigcup \; \{ \; (c,r) \; | \;$ \\
	\> ~ $(\exists\, q,c,r: \; q\in N_p: \; c=\cache\textit{color}_q
	\;\wedge\; r=\cache\textit{min}\ell_q) \, \}$
\end{tabbing}
\begin{lemma} \label{color-time}
The algorithm \textsf{R4}-\textsf{R8} converges
to a distance-two coloring, with probability 1;  the  
convergence time is $O(1)$ if each timed 
variable propagation completes in $O(1)$ time.
\end{lemma}

\begin{proof}
The proof is a sequence of observations to reflect the essentially
sequential character of color assignment.  We consider an execution
where the set of leaders has been established by \textsf{R1}--\textsf{R3}
initially.   Observe that in
$O(1)$ time the assignments of \textsf{R6} reach a fixed point,
based on the local reference to $\cache\ell_q$ for neighbors.  Therefore,
in $O(1)$ time, the shared variables $min\ell_p$ are propagated to 
$N_p$ and caches $\cache min\ell_p$ are stable.
Similarly, in $O(1)$ additional time, the assignments of \textsf{R4} 
reach a fixed point, so that leaders have stable \textit{dom} variables.
 
The remainder of the proof is an induction to show that color assignments
stabilize in $O(\Delta)$ phases 
(recall that $\Delta$ is the constant of Lemma \ref{dag-lemma}).
For the base case of the induction, consider the set $S$ 
of leader nodes such that for every $p\in S$, within $N^3_p$ 
no leader of smaller name than $p$ occurs.
We use distance three rather than distance two so that such a leader 
node's choice of colors is stable, independent of the choices made by
other leaders.  Set $S$ is non-empty because, of the set of leaders
in the network, at least one has minimal name, which is unique up to 
distance three.  Call $S$ the set of \emph{root} leaders.
Given such a leader node $p$, 
each neighbor $q\in N_p$ executes \textsf{R8} within
$O(1)$ time and assigns to $\textit{spectrum}_q$ a set of tuples with the
property that for any $(c,r)\in\textit{spectrum}_q$, $r\geq p$.  
Notice that although $\textit{spectrum}_q$ could subsequently change
in the course of the execution, this property is stable.  Therefore,
in $O(1)$ additional time, no tuple of $\cache\textit{spectrum}_q$ 
has a smaller value than $p$ in its second component.  It follows that
any subsequent evaluation of \textsf{R5} by leader $p$ has a fixed 
point:  $p$ assigns colors to all nodes of $N_p$.  After $O(1)$
delay, for $q\in N_p$, $\cache\textit{setcol}_p$ stabilizes.  Then in $O(1)$
time, all nodes of $\textit{dom}_p$ assign their \textit{color} variables
using \textsf{R7}.  This completes the base case, 
assignment of colors by root leaders. 
 
We complete the induction by examining nodes with minimum distance $k>0$ from
any root leader along a path of increasing leader names (referring to the
\textit{Up} predicate used in Lemma \ref{dag-lemma}).  The hypothesis for
the induction is that nodes up to distance $k-1$ along an increase path
of leader names have stabilized to a permanent assignment of colors to the 
nodes they dominate.  Arguments similar to the base case show that 
such nodes at distance $k$ eliminate colors already claimed by leaders  
of the hypothesis set in their evaluations of \textsf{R5}.  The entire
inductive step --- extending by one all paths of increasing names from
the root leaders --- consumes $O(1)$ additional time. 
The induction terminates at $\Delta$, thanks to Lemma \ref{dag-lemma}, hence
the overall bound of $O(\Delta)$ holds for convergence.
\end{proof}

Only at one point in the proof do we mention distance-three
information, which is to establish the base case for root leaders  
(implicitly it is also used in the inductive step as well).
Had we only used neighborhood naming unique up to distance two, it 
would not be ensured that a clear ordering of colors exists between 
leaders that compete for dominated nodes, \emph{eg}, a leader $p$ could
find that some node $r\in N^2_p$ has been assigned a color by another 
leader $q$, but the names of $p$ and $q$ are equal; this conflict 
would permit $q$ to assign the same color to $r$ 
that $p$ assigns to some neighbor of $r$.  We use distance-three
unique naming to simplify the presentation, rather than presenting a 
more complicated technique to break ties.  Another useful intuition 
for an improved algorithm is that Lemma \ref{dag-lemma}'s result is 
possibly stronger than necessary:  if paths of increasing names  
have at most some constant length $d$ with high probability, and
the algorithms for leader selection and color assignment tolerate
rare cases of naming conflicts, the expected convergence time would
remain $O(1)$ in the construction.

Due to space restrictions, we omit the proof that the resulting
coloring is minimal (which follows from the construction of $f$ to
be locally minimum, and the essentially sequential assignment of 
colors along paths of increasing names).

\section{Assigning Time Slots from Colors} \label{sec:slots}

Given a distance-two coloring of the network nodes, the 
next task is to derive time slot assignments for each node
for TDMA scheduling.  Our starting assumption is that each
node has equal priority for assigning time slots, \emph{ie},
we are using an unweighted model in allocating bandwidth.
Before presenting an algorithm, we have two motivating observations.  

First, the algorithms that
provide coloring are local in the sense that the actual 
number of colors assigned is not available in any global 
variable.  Therefore to assign time slots consistently to 
all nodes apparently requires some additional computation.
In the first solution of Figure \ref{fig:solution-A}, 
both leftmost and rightmost nodes have color 1, however
only at the leftmost node is it clear that color 1 should be  
allocated one ninth of the time slots.  Local information 
available at the rightmost node might imply that color 1 
should have one third of the allocated slots.  

The second observation is that each node $p$ should have about 
as much bandwidth as any other node in $N^2_p$.   This follows
from our assumption that all nodes have equal priority.  Consider  
the $N^2_p$ sizes shown in Figure \ref{fig:solution-B} that 
correspond to the colorings of \ref{fig:solution-A}.  The rightmost
node $p$ in the first coloring has three colors in its two-neighborhood,
but has a neighbor $q$ with four colors in its two-neighborhood.  It 
follows that $q$ shares bandwidth with four nodes: $q$'s share of
the bandwidth is at most 1/4, whereas $p$'s share is at most
1/3.  It does not violate fairness to allow $p$ to use 1/3 of
the slot allocation if these slots would otherwise be wasted. 
Our algorithm therefore allocates slots in order from most 
constrained (least bandwidth share) to least constrained, so that 
extra slots can be used where available.

To describe the algorithm that allocates media access time for 
node $p$, we introduce these shared variables and local functions.
\begin{description}
\item[$base_p$] stabilizes to the number of colors in $N^2_p$.
The value $base_p^{-1}=1/base_p$ is used as a constraint on the share
of bandwidth required by $p$ in the TDMA slot assignment.
\item[$itvl_p$] is a set of intervals of the form 
$[x,y)$ where $0\leq x<y\leq 1$.  For allocation, each unit of time
is divided into intervals and $\textit{itvl}_p$ is the set of intervals
that node $p$ can use to transmit messages.  The expression 
$|[x,y)|$ denotes the time-length of an interval.
\item[$g(b,S)$] is a function to assign intervals, where $S$ is a set of 
subintervals of $[0,1)$.  Function $g(b,S)$ returns a maximal set $T$ of 
subintervals of $[0,1)$ that are disjoint and also disjoint 
from any element of $S$ such that $(\sum_{a\in T} |a|) \leq b$.
\end{description}
To simplify the presentation, we introduce $S_p$ as a private (nonshared)
variable.
\begin{tabbing}
xxx \= xxxxx \= xxx \= \kill
\>\textsf{R9:} \> $\textit{true} \;\rightarrow\; 
	\textit{base}_p := |\; \{\,\cache\textit{color}_q\; 
		| \; q\in N^2_p \,\}\;|$ \\
\>\textsf{R10:} \> $\textit{true} \;\rightarrow\;  
	S_p := \bigcup \; \{ \, \cache\textit{itvl}_q \; | \;  
	q\in N^2_p \quad\wedge$ \\
\>\> $(\cache\textit{base}_q > \textit{base}_p \;\vee\;$ \\
\>\> $(\cache\textit{base}_q=\textit{base}_p \;\wedge\;
		\cache\textit{color}_q < \textit{color}_p)) \; \}$ \\
\>\textsf{R11:} \> $\textit{true} \;\rightarrow\;  
	\textit{itvl}_p := g(\textit{base}_p^{-1},S_p)$ 
\end{tabbing}
\begin{lemma}
With probability 1 the algorithm \textsf{R9--R11} 
converges to an allocation of time intervals such that no two
nodes within distance two have conflicting time intervals, and 
the interval lengths for each node $p$ sum to 
$|\{\,\textit{color}_q\;|\; q\in N^2_p\,\}|^{-1}$;  the
expected convergence time of \textsf{R9}-\textsf{R11} is 
$O(1)$ starting from any state with stable and valid coloring.
\end{lemma}
\begin{proof}
Similar to that for Lemma \ref{color-time}; we omit details.
\end{proof}
It can be verified of \textsf{R9-R11} that, at a fixed
point, no node $q\in N^2_p$ is assigned a time that
overlaps with interval(s) assigned to $p$;  also,
all available time is assigned (there are no leftover intervals). 
A remaining practical issue is the conversion
from intervals to a time slot schedule:  a discrete TDMA slot schedule
will approximate the intervals calculated by $g$.

\section{Assembly} \label{sec:assembly}

Given the component algorithms of Sections  
\ref{sec:identification}--\ref{sec:slots}, the 
concluding statement of our result follows.

\begin{theorem}
The composition of \textsf{N0}--\textsf{N3} and 
\textsf{R1}--\textsf{R11} is a probabilistically 
self-stabilizing solution to $\mathcal{T}$ with
$O(1)$ expected local convergence time.
\end{theorem}
\begin{proof}
The infrastructure for neighborhood discovery and 
shared variable propagation \textsf{N0}--\textsf{N2} contributes $O(1)$ 
delay (partly by assumption on the CSMA/CA behavior), and \textsf{N3} 
establishes neighborhood unique naming in expected $O(1)$ time.   The
subsequent layers \textsf{R1--R3}, \textsf{R4--R8}, and \textsf{R9--R11},
each have $O(1)$ expected convergence time, and each layer is only 
dependent on the output of the previous layer.  The hierarchical 
composition theorem (see \cite{Tel94}) implies overall stabilization,
and the expected convergence time is the sum of the expected convergence
times of the components.
\end{proof}

\section{Conclusion} \label{sec:conclusion}

Sensor networks differ in characteristics and in typical applications
from other large scale networks such as the Internet.  Sensor networks
of extreme scale (hundreds of thousands to millions of nodes) have been
imagined \cite{KKP99}, motivating scalability concerns for such
networks.  The current generation of sensor networks emphasizes the 
\emph{sensing} aspect of the nodes, so services that aggregate data 
and report data have been emphasized.  Future generations of sensor 
networks will have significant actuation capabilities.  In the context
of large scale sensor/actuator networks, end-to-end services can be 
less important than regional and local services.  Therefore we 
emphasize local stabilization time rather than global stabilization 
time in this paper, as the local stabilization time is likely to be
more important for scalability of TDMA than global stabilization time.
Nonetheless, the question of global stabilization time is neglected 
in previous sections.  We speculate that global stabilization time 
will be sublinear in the diameter of the network (which could be a
different type of argument for scalability of our constructions, 
considering that end-to-end latency would be linear in the network
diameter even after stabilization).  Some justification for our 
speculation is the following:  if the expected local time for 
convergence is $O(1)$ and underlying probability assumptions 
are derived from Bernoulli (random name selection) and Poisson
(wireless CSMA/CA) distributions, then these distributions can be 
approximately bounded by exponential distributions with constant
means.  Exponential distributions define half-lives for populations
of convergent processes (given asymptotically large populations), 
which is to say that within some constant time $\gamma$, the 
expected population of processes that have not converged is halved;
it would follow that global convergence is $O(\lg n)$.  

We close by mentioning two important open problems.  Because sensor
networks can be deployed in an \emph{ad hoc} manner, new sensor 
nodes can be dynamically thrown into a network, and mobility is 
also possible, the TDMA algorithm we propose could have a serious
disadvantage:  introduction of just one new node could disrupt 
the TDMA schedules of a sizable part of a network before the 
system stabilizes.  Even if the stabilization time is expected
to be $O(1)$, it may be that better algorithms could contain  
the effects of small topology changes with less impact than our
proposed construction.  One can exploit normal notifications
of topology change as suggested in \cite{DH97}, for example.
Another interesting question is whether 
the assumption of globally synchronized clocks (often casually 
defended by citing GPS availability in literature of wireless
networks) is really needed for self-stabilizing TDMA construction;
we have no proof at present that global synchronization is 
necessary.

%end small

\end{document}